\documentclass[epjCONF, onecolumn]{svjour}
\usepackage{graphics}
\usepackage[varg]{txfonts} % Times fonts
\usepackage[latin1]{inputenc}

\def\rhodm{\rho_\mathrm{dm}}
\def\rhos{\rho_\mathrm{s}}
\def\msun{\,M$_\odot$}

\session-title{Assembling the puzzle of the Milky Way}
\begin{document}
\title{Limits on the local dark matter density}
\author{Silvia Garbari\inst{1}\fnmsep\thanks{\email{silvia@physik.uzh.ch}} \and Justin I. Read\inst{2,3} \and George Lake\inst{1}}
\institute{Institute of Theoretical Physics, University of Z\"urich \and 
Department of Physics \& Astronomy, University of Leicester \and Institute for Astronomy, Department of Physics, ETH Z\"urich}
\abstract{We study the systematic problems in determining the local dark matter density $\rhodm(R_\odot)$ from kinematics of stars in the Solar Neighbourhood, using a simulated Milky Way-like galaxy. We introduce a new unbiased method for recovering $\rhodm(R_\odot)$ based on the moments of the Jeans equations, combined with a Monte Carlo Markov Chain (MCMC) technique and apply it to real data \cite{garbari}.}
\maketitle

\section{Introduction}
A good constraint on the local dark matter (DM) density gives us useful information about local Galaxy dynamics. It is also vital for direct DM search experiments. With near-future surveys promising a dramatic improvement in the number and precision of astrometric, photometric and spectroscopic local stars data, it is timely to revisit the systematic problems arising in determining the local matter density, as well as the local dark matter density - $\rhodm(R_\odot)$ - from the vertical motion of stars at the Solar Neighbourhood. To do this, we use a high resolution N-body simulation of a Milky Way-like galaxy to test the standard methods in the literature. We also introduce a new method for recovering $\rhodm(R_\odot)$ based on the moments of the Jeans equations, combined with a MCMC technique to marginalise over the unknown parameters. This method relies on very few assumptions, but requires more precise data than the classical methods adopted in the literature. Given sufficiently good data, our method recovers the correct local dark matter density even in the face of disc inhomogeneities, non-isothermal tracers and a non-separable distribution function \cite{garbari}.
 %end of abstract
%

%
%\section{Introduction}
%\label{intro}

\section{Method and results}\label{sec:1}
The standard methods to determine the local density from the kinematics of stars \cite{hf2000,hf2004,kg1989,bahcall} rely on several assumptions:
\begin{enumerate}
\item The disc is in equilibrium (steady state assumption) \cite{hf2000,hf2004,kg1989,bahcall}; \label{1}
\item The local dark matter density is a constant $\rhodm(R_\odot)$ in the $z-$range considered  ($\lesssim$1\,kpc) \cite{hf2000,hf2004,bahcall};\label{2}
\item Tilt terms involving $\sigma_{zR}$ in the Jeans equation are negligible \cite{hf2000,hf2004,kg1989,bahcall};\label{3}
\item The local visible matter is a sum of isothermal components \cite{hf2000,hf2004,bahcall};\label{4}
\item The rotation curve of the Milky Way (MW) is flat \cite{hf2000,hf2004,kg1989,bahcall};\label{5}
\item The distribution function of the tracers is a function only of the vertical energy: $f=f(E_z)$ \cite{hf2000,hf2004,bahcall}. In this case the density law of a stellar tracer $\nu(z)$ can be predicted from its vertical velocity distribution at the plane $f(v_{z0})$, in a given potential $\Phi(z)$\label{6}\footnote{Actually, this implies that $\sigma_{zR}=0$, rhather than just very small.}:
\begin{center}
$\nu(z)=2\int_{\sqrt{2\Phi}}^\infty\frac{f(v_{z0})v_{z0}dv_{z0}}{\sqrt{v_{z0}^2-2\Phi(z)}}$\end{center}
\end{enumerate}
The comparison between the predicted and the observed $\nu(z)$ of the tracer stars constrains the local potential, and thus the local visible and dark matter densities \cite{hf2000,hf2004,bahcall}.
Our new method, based on the solution of the Jeans-Poisson system, (i) relies on a Minimal set of Assumptions ({\it MA method} \cite{garbari}) -- namely \ref{1}, \ref{2}, \ref{3} -- and (ii) uses a MCMC technique to marginalise over unknown parameters. The former makes the method -- given good enough data -- robust to model systematics; the latter allows us to cope with incomplete or noisy data and model degeneracies. In the MA method the density law of a tracer population $\nu(z)$ - in a given potential $\Phi(z)$ - is predicted from the solution of the Jeans equation
\begin{center}
 $\frac{\nu(z)}{\nu(0)}=\frac{\sigma_z^2(0)}{\sigma_z^2(z)}\exp\left(-\int_0^z\frac{1}{\sigma_z^2}\frac{d\Phi}{dz}dz\right)$\end{center}
where $\sigma_z^2(z)$ is the vertical velocity dispersion law.

\begin{figure}
\centering
\resizebox{0.75\columnwidth}{!}{%
\includegraphics{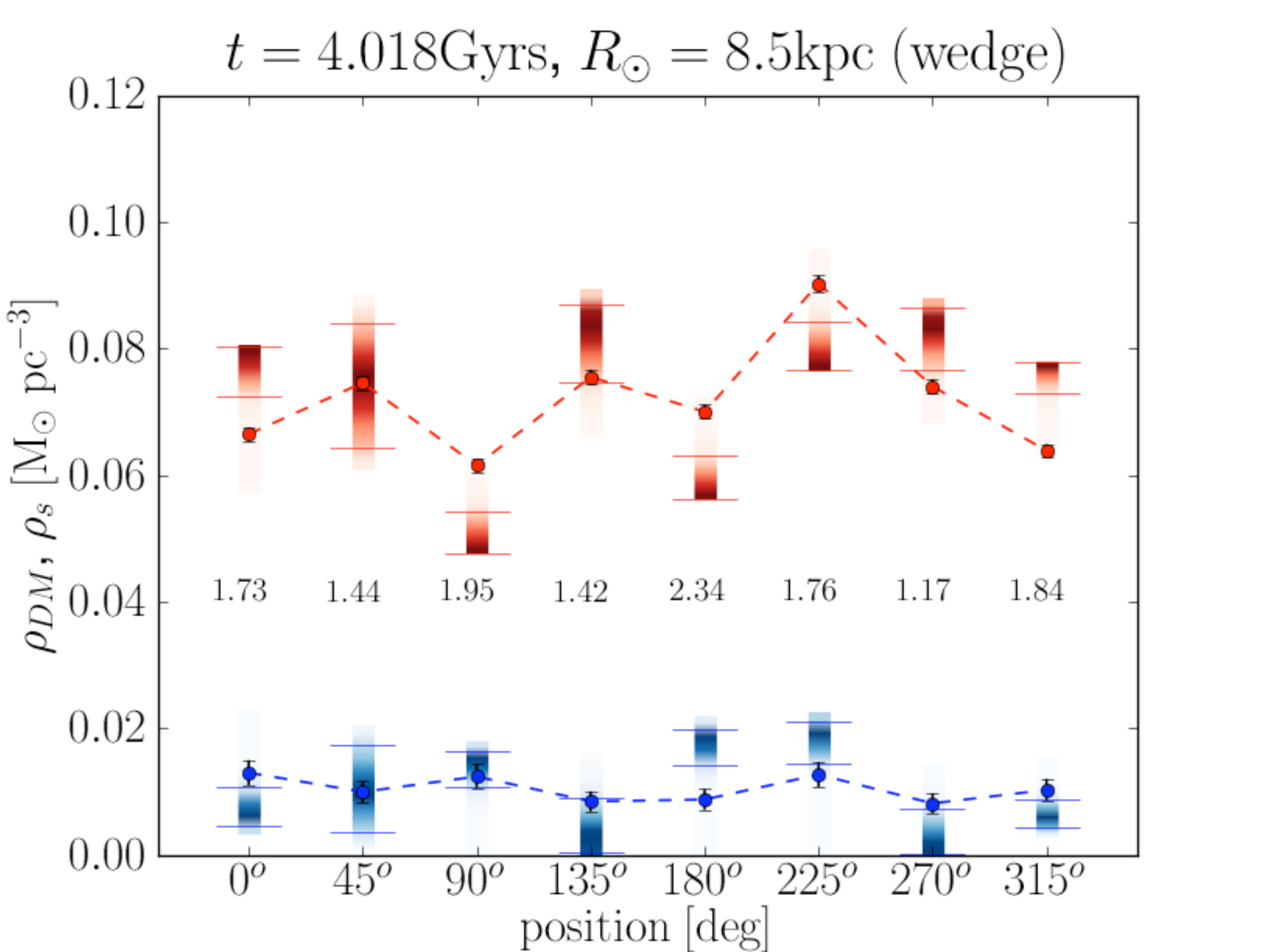}
\includegraphics{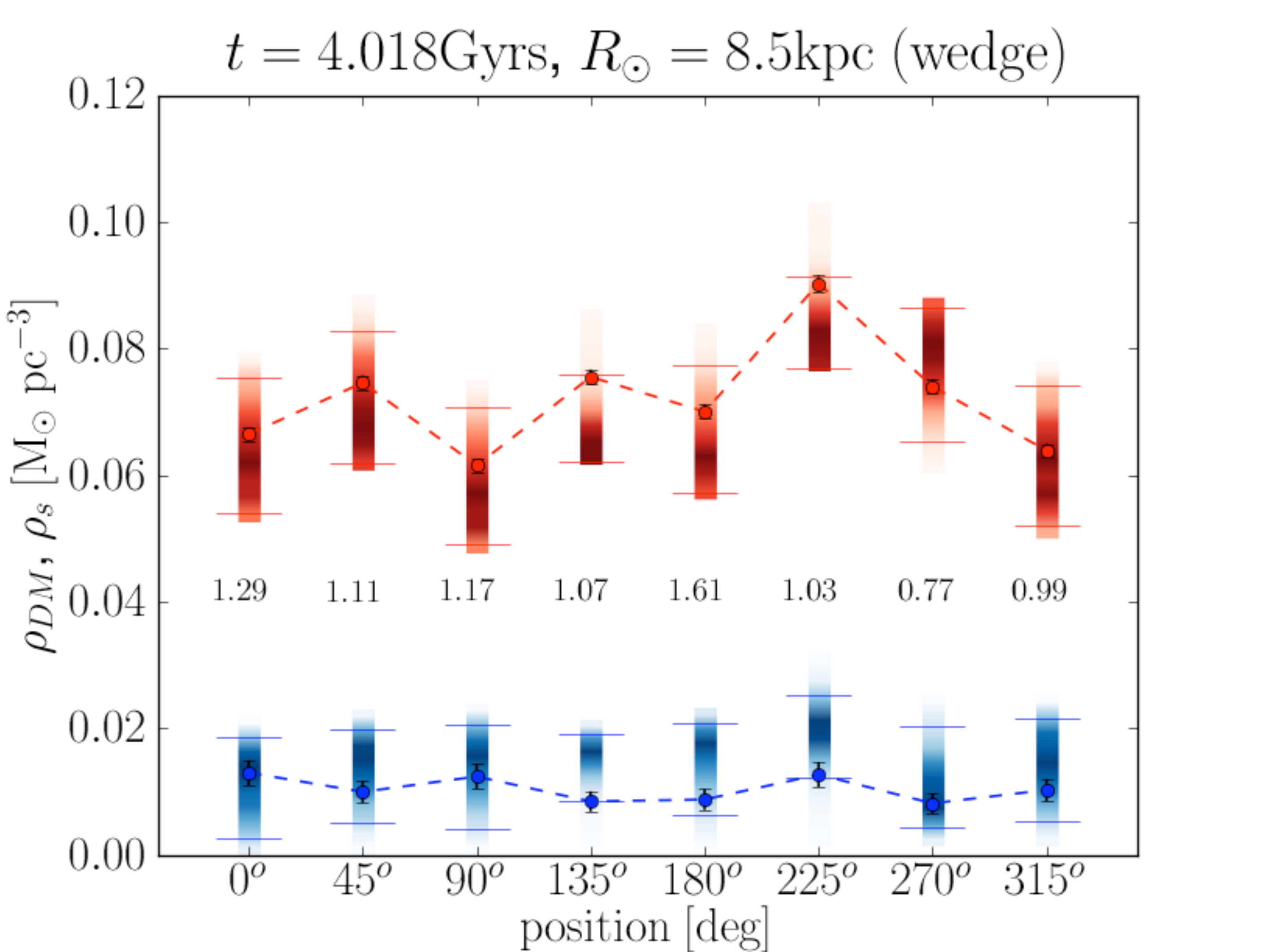}
}
\caption{Density of models explored by the MCMC (for 8 volumes around the disc) represented as shaded areas of different colours (blue=$\rhodm$, red=$\rhos$). The filled dots represent the corresponding actual values from the simulation (with Poisson error bars), the horizontal segments are the 90\% confindence ranges. Left panel: standard method, Right Panel: MA method.}
\label{fig:1}       % Give a unique label
\end{figure}

%\section{Test of the methods using the simulation}
To test both the standard and the MA method, we build a high resolution equilibrium N-body model approximating the Milky Way that satisfies all of the usual assumptions. We then evolve the disc over $\sim 4$\,Gyr to form a complex disc structure with a strong bar and spiral waves similar to the MW. We fit $\nu(z)$ to find the best value of of the local visible and dark matter densities $\rhos$ and $\rhodm$ at 8 different angular positions around the disc, representing the Solar Neighbourhood (at $R=8.5$\,kpc from the Galactic centre).
The application of the standard and the MA methods to the unevolved simulation gives similar results and recovers the correct value of $\rhos$ and $\rhodm$. However, we find that we need data extending up to $\sim3$ times the half mass scale height of the disc to break a very strong degeneracy between $\rhos$ and $\rhodm$ (see also \cite{bahcall}).
The situation is very different for the evolved disc that does not satisfy assumptions \ref{4}, \ref{5} and \ref{6}. While it is easy to correct for \ref{4} and \ref{5}, the failure of hypothesis \ref{6} makes the standard methods fail in recovering the correct $\rhos$ and $\rhodm$ (see fig. \ref{fig:1}).

%\section{Application to real data}
We illustrate our MA method by applying it to Hipparcos data \cite{hf2000,hf2004}, extending  to $\sim 700$\,pc. We first make the usual assumption that the (A, F and K giant) star tracer populations are isothermal. This recovers $\rhodm=0.003^{+0.009}_{-0.007}$\msun\,pc$^{-3}$  (with 90\% confidence), consistent with previous determinations. However, the vertical dispersion profile of these tracers is poorly known. If we assume instead a non-isothermal profile similar to the blue disc stars from SDSS DR-7 \cite{bond}, we obtain an equally good fit to the tracer density profile, but with ut with $\rhodm=0.033^{+0.008}_{-0.009}$\msun\,pc$^{-3}$. This highlights the fact that it is vital to measure the vertical dispersion profile of the tracers to recover an unbiased estimate of $\rhodm$. 

%\section{Conclusions}
%\begin{enumerate}
%\item Data up to significantly larger z than the MW disc scale height are required to break a degeneracy between $\rhodm$ and the local visible matter density $\rhos$. 

%\item Methods assuming that the distribution function of the tracers is a function only of the vertical energy $f=f(E_z)$ become systematically biased if the motion is not truly separable in $z$. This effect becomes important when fitting to data that extend to large $z$ (which is necessary to break the $\rhodm-\rhos$ degeneracy).

%\item We introduce a new minimal assumption (MA) method that correctly recovers both $\rhodm$ and $\rhos$ even in the face of disc inhomogeneities, non-separability of the $z-$motion, and vertical non-isothermality of the tracers, provided that $\sigma_z(z)$ is known. 
%\end{enumerate}

%and \cite{RefJ}
%\subsection{Subsection title}
%\label{sec:2}
%as required. Don't forget to give each section
%and subsection a unique label (see Sect.~\ref{sec:1}).
%

%\begin{table}
%\caption{Please write your table caption here.}
%\begin{tabular}{lll}
%\hline\noalign{\smallskip}
%\end{tabular}
%\end{table}
%

\end{document}